\documentclass[aps,prd,
showpage,nobibnotes,groupaddress,amssymb,amsmath,nofootinbib,floatfix]{revtex4-2}

\usepackage{graphicx,amsfonts,amssymb}
\usepackage{amsmath}
\usepackage{epsfig}
\usepackage{dcolumn}
\usepackage{bm}
\usepackage{siunitx}
\usepackage[hidelinks]{hyperref}
\allowdisplaybreaks

\begin{document}

\title{An exact lower bound within a 331 model closing solution}


\author{Mario W. Barela}
\email{mario.barela@unesp.br}
\affiliation{
Instituto de F\'\i sica Te\'orica, Universidade Estadual Paulista, \\
R. Dr. Bento Teobaldo Ferraz 271, Barra Funda\\ S\~ao Paulo - SP, 01140-070,
Brazil}

\begin{abstract}%
We show how a specific proposal of solution to the problem of `closing' the well motivated 331 model to the Standard Model actually implies a lower bound for the otherwise theoretically free vacuum expectation value $v_\chi$. 
\end{abstract}

\maketitle

\section{Introduction}
\label{sec:intro}

Besides a gauge theoretically sensible construction, an extension of the Standard Model should match its phenomenological predictions numerically at a well known scale. In practice, this means that, at that scale, we should be able to write at least some free parameters of the new model algebraically in terms of Standard Model ones, tuned in such a way that measured quantities like charged or neutral current parameters and known particle masses are reproduced within the extension.

One such `solution' to close the new physics model symmetry to that of the Standard Model at the scale of the $Z$ boson mass was proposed for the 331 model \cite{Dias:2006ns}. For its minimal version (already possibly capable of reproducing the neutrino masses and removing the arbitrariness in the number of families)\cite{Pisano:1991ee, Frampton:1992wt, Foot:1992rh}, a version with exotic heavy leptons \cite{Pleitez:1992xh} and one with right-handed neutrinos \cite{Dias:2005yh}, Dias et al found a remarkably simple relation among parameters that reduces known boson masses to the expected ones. The authors have also verified that imposing the solution fixes the neutral current parameters to their SM values.  

In this note, for the minimal model and the model with right handed neutrinos\footnote{The model with heavy leptons is identical, in the subject of this study, to the minimal model.}, it is shown that the reduction to the SM expressions is actually not guaranteed by the solution alone, and it is the needed extra condition that, through an expression form branching resulting from trivial algebraic manipulations, sets a lower bound on the $\chi$ scalar triplet vacuum expectation value, which is responsible for the breaking $SU(3)_C\otimes SU(3)_L \otimes U(1)_X \to \text{SM}$. We also derive and catalog again all the relevant expressions for general neutral current parameters and masses within these models' general and simplified (constrained by the solution) versions, many of which differ from those presented in \cite{Dias:2006ns}.

\section{Minimal model bounds}
\label{sec:minimal}

\subsection{Model review}
\label{sec:minreview}

This description is only intended to discern explicitly the fermionic (whose neutral current parameters we calculate) and scalar (as to define the vacuum expectation values of the theory) representation content. For a complete assessment of the model, we refer to \cite{Pisano:1991ee,Frampton:1992wt,Foot:1992rh}.

The leptonic sector of the minimal model is composed of a left-handed triplet $\allowbreak (\nu_a,\ell_a,\ell_a^c)_L^T \sim (\mathbf{1},\mathbf{3},0)$ and a right-handed neutrino $\nu_R \sim (\mathbf{1},\mathbf{1},0)$, which could possibly be omitted. The quark representation content is asymmetric, with two anti-triplets $(d_m,u_m,j_m)_L^T \sim (\mathbf{3},\mathbf{3}^*,-1/3)$ and a triplet $(u_3,d_3,J)_L^T \sim (\mathbf{3},\mathbf{3},2/3)$, besides the right-handed representations $u_{\alpha R} \sim (\mathbf{3},\mathbf{1},2/3)$, $d_{\alpha R} \sim (\mathbf{3},\mathbf{1},-1/3)$, $J_{R} \sim (\mathbf{3},\mathbf{1},5/3)$, $j_{m R} \sim (\mathbf{3},\mathbf{1},-4/3)$, where $m=1,2$ and $\alpha=1,2,3$. This non-democratic property of the matter content is what enforces anomaly cancellation, and effectively fixates the number of families to three (in principle, any multiple of three could be possible, but a number larger than three is ruled out by the QCD asymptotic freedom fitting).

{\sloppy The scalar sector accommodates 3 triplets: $(\eta^0,\eta_1^-,\eta_2^+)^T \sim (\mathbf{1},\mathbf{3},0)$, $\allowbreak(\rho^+, \rho^0 , \rho^{++})^T \sim (\mathbf{1},\mathbf{3},1)$, $(\chi^-,\chi^{--},\chi^{0})^T \sim (\mathbf{1},\mathbf{3},-1)$ and a sextet}

\begin{equation}
\begin{pmatrix}
\sigma_1^0 & \frac{h_2^+}{\sqrt{2}} & \frac{h_1^-}{\sqrt{2}} \\
\frac{h_2^+}{\sqrt{2}} & H_1^{++} & \frac{\sigma_2^0}{\sqrt{2}} \\
\frac{h_1^-}{\sqrt{2}} & \frac{\sigma_2^0}{\sqrt{2}} & H_2^{--}
\end{pmatrix} \sim (\mathbf{1},\mathbf{6}^*,0).
\end{equation}
The VEVs are denoted $\langle \eta^0 \rangle = \frac{v_\eta}{\sqrt{2}}$, $\langle \rho^0 \rangle = \frac{v_\rho}{\sqrt{2}}$, $\langle \chi^0 \rangle = \frac{v_\chi}{\sqrt{2}}$ and $\langle \sigma_2^0 \rangle = v_{s}$, $\langle \sigma_1^0 \rangle =0$.

The exact masses of the neutral vector bosons are calculated to be 

\begin{equation}\label{eq:minMZ1}
M_{Z_1}=\frac{g^2}{6}\left[v_W^2+3t^2 v_\rho^2 +v_\chi^2+3t^2 v_\chi^2 -R     \right],
\end{equation}
\begin{equation}
M_{Z_2}=\frac{g^2}{6}\left[v_W^2+3t^2 v_\rho^2 +v_\chi^2+3t^2 v_\chi^2 -R      \right],
\end{equation}
where
 
\begin{equation}
R \equiv \left\{ v_W^4 - v_W^2 \left[ (3+6t^2)v_\rho^2 + (1+6t^2)v_\chi^2 \right] + (1+ 3t^2 )\left[ 3(1+t^2)v_\rho^2 + 6t^2 v_\rho^2 v_\chi^2 +  (1+ 3t^2 ) v_\chi^4 \right]\right\}^{1/2}.
\end{equation}

We will soon make use of the fact that the exact mass of the $W$-boson mass in the present model, and also in the model which will be examined later, is 

\begin{equation}\label{eq:MW}
M_W^2=g_{3L}^2\frac{v_\eta^2+v_\rho^2+2v_s^2}{4},
\end{equation}
where $g_{3L}$ is the coupling constant relative to $SU(3)_L$. This is used to, comparing with the SM, eliminate one of the previously free VEVs

\begin{equation}\label{eq:veta}
v_\eta^2+v_\rho^2+2v_s^2 = v_W^2 \implies v_\eta=\sqrt{v_W^2 - v_\rho^2 - 2v_s^2}.
\end{equation} 

\subsection{Solution}

In order to guarantee the closing of the 331 symmetry at the electroweak scale -- \textit{i.e.}, that all precisely known numerical predictions of the standard model are reproduced, specially the $Z$-boson mass -- the first obvious possibility would be to make $v_\chi$ formally infinite. In practice, this renders the higher symmetry broken from the start. We call this the trivial solution to the closing.

A much more interesting non-trivial solution \cite{Dias:2006ns} is defined by imposing, on the electroweak scale, the relation

\begin{equation} \label{eq:soldef}
v_\rho^2=\frac{1-4s_W^2}{2c_W^2}v_W^2,
\end{equation}
where the quantities appearing here are the sine and cosine of the weak angle and the known Higgs expectation value, such that $M_Z=M_W/c_W=gv_W/2c_W$.

It also needs to be enforced that

\begin{equation} \label{eq:gequal}
g_{3L}=g_{2L},
\end{equation}
which, on Renormalization Group Equation terms, is a matching condition consequent of the embedding structure $SU(2)_L \subset SU(3)_L$, necessarily valid at a single scale identified with the breaking $331 \to \text{SM}$ and should be set as a boundary condition when one calculates full model runnings. It is usually and in this case assumed to be a good approximation at the electroweak scale. 

Together with the SM relation $1/e^2=1/g_{SU(2)}^2+1/g_{Y}^2$, this allows us to write $t$ in terms of the SM electroweak angle. In the minimal model this relation reads

\begin{equation} \label{eq:tXsW}
t^2 \equiv \frac{g_X^2}{g^2} = \frac{s_W^2}{1-4 s_W^2}.
\end{equation}   
 \cite{Dias:2006ns} claims that, using the above relation, Eq.~(\ref{eq:soldef}) ensures that known boson masses within the 331 structure reproduce the correct values of the SM. The authors also verify that the solution is stable from a custodial symmetry perspective.

\subsection{Bounds}

The mechanism that gives rise to the bounds is trivial. Consider the general $Z$-boson mass given by Eq. (\ref{eq:minMZ1}). To obtain the mass on the solution-constrained model, we plug into it Eqs. (\ref{eq:soldef}-\ref{eq:tXsW}), with which the $R$ term simplifies to 

\begin{equation}
R = \sqrt{\left[ (-1+4 s_W^2)v_W^2 + 2 c_W^2 v_\chi^2 \right]^2}.
\end{equation}
This causes the solution-constrained form for this mass to branch like 

\begin{equation}
M_Z^2=\begin{cases}
\frac{gv_W^2}{4c_W^2}, & \text{if} \ v_\chi \geq \sqrt{\frac{1-4s_W^2}{2 c_W^2}}v_W \\
\frac{g^2[(1-2s_W^2-8s_W^4)v_W^2+4c_W^4v_\chi^2]}{12(1-5s_W^2+4s_W^4)},  & \text{otherwise}
\end{cases}
\end{equation}
of which the first part, the true SM $Z$-boson mass, must be chosen.


The same sort of analysis needs to be effected on the neutral current parameters, whose general exact forms, useful for the model's phenomenology, are presented on \ref{app:couplingsminimal}. It happens that the very condition above also selects the correct branch for all of the known couplings (see \ref{app:couplingssolminimal}) and reads 

\begin{equation} \label{eq:bound}
v_\chi \geq \sqrt{\frac{1-4s_W^2}{2 c_W^2}}v_W \approx \SI{54}{GeV},
\end{equation} 
where we have used the values $v_W=\SI{246.22}{GeV}$ and $s_W^2=0.23121$\cite{Zyla:2020zbs}. 

This indicates that for the ansatz proposed in \cite{Dias:2006ns} to represent a true solution to the closing, Eq. (\ref{eq:bound}) needs to be imposed besides the original definition of Eq. (\ref{eq:soldef}). In this way, all required physical parameters are reproduced in the referred scale.

This, in turn, means that the mass of the doubly-charged vector boson dilepton obeys

\begin{equation}
M_U=\sqrt{\frac{1}{4}g^2(v_\rho^2+2v_s^2+v_\chi^2)} \geq \SI{24}{GeV}.
\end{equation}

This lower bound is actually slightly lifted by the still free $v_s$. $v_s$, however, should not be greater than a few GeV once it is responsible for fitting neutrino masses, rendering it relatively negligible, which ensures that the above is an almost maximal lower bound for the $U^{++}$ mass.

Now, noting that in the solution $v_\rho$ is numerically defined, we may evoke Eq. (\ref{eq:veta}) to find a lower bound for the singly-charged vector dilepton $V^+$

\begin{equation}
M_V=\sqrt{\frac{1}{4}g^2(v_\eta^2+2v_s^2+v_\chi^2)} \geq \SI{78}{GeV}.
\end{equation}

Lastly, the mass of the exotic neutral vector boson which, in the solution, reads

\begin{equation}
M_{Z'}=\frac{g \,v_\chi^2}{12}\:\frac{4 c_W^2 + (1-2s_W^2-8s_W^4)\bar{v}_W^2}{c_W^2 (1-4 s_W^2)},
\end{equation}
(where $\bar{v}_W=v_W/v_\chi$), is also bounded:

\begin{equation}
M_{Z'}\geq \SI{89}{GeV}.
\end{equation}

\section{Model with right-handed neutrinos}
\label{sec:righthanded}

\subsection{Model review}
\label{sec:rhreview}

For an assessment of the model, see \cite{Dias:2005yh}.

\sloppy The leptonic representation content now consists of $(\nu_a,\ell_a,\nu_a^c)_L^T \sim (\mathbf{1},\mathbf{3},-1/3)$, $e_R \sim (\mathbf{1},\mathbf{1},-1)$. The quark sector is composed of $(d_m,u_m,D_m)_L^T \sim (\mathbf{3},\mathbf{3}^*,0)$, $(u_3,d_3,U)_L^T \sim (\mathbf{3},\mathbf{3},1/3)$, $u_{\alpha R} \sim (\mathbf{3},\mathbf{1},2/3)$, $d_{\alpha R} \sim (\mathbf{3},\mathbf{1},-1/3)$, $U_{R} \sim (\mathbf{3},\mathbf{1},2/3)$, $D_{m R} \sim (\mathbf{3},\mathbf{1},-1/3)$.

\sloppy The scalar sector contains: $(\eta^0,\eta^-,\eta^{\prime 0})^T \sim (\mathbf{1},\mathbf{3},-1/3)$, $(\rho^+,\rho^0,\rho^{\prime +})^T \sim (\mathbf{1},\mathbf{3},2/3)$, $(\chi^0,\chi^{-},\chi^{\prime 0})^T \sim (\mathbf{1},\mathbf{3},-1/3)$. The nonzero VEVs are $\langle \eta^0 \rangle = \frac{v_\eta}{\sqrt{2}}$, $\langle \rho^0 \rangle = \frac{v_\rho}{\sqrt{2}}$, $\langle \chi^{\prime 0} \rangle = \frac{v_{\chi^\prime}}{\sqrt{2}}$.

The neutral vector boson masses are given by

\begin{equation}\label{eq:rhnMZ1}
M_{Z_1}=\frac{g^2}{18}\left[ (3+t^2)v_W^2+3t^2 v_\rho^2 +3v_\chi^2+t^2 v_\chi^2 -R     \right],
\end{equation}
\begin{equation}
M_{Z_2}=\frac{g^2}{18}\left[ (3+t^2)v_W^2+3t^2 v_\rho^2 +3v_\chi^2+t^2 v_\chi^2 +R     \right],
\end{equation}
where

\begin{equation}
\begin{split}
R \equiv \left\{ (3+t^2) \left[(3+t^2) (v_W^2 \right. \right. & \left. + v_\chi^2)^2 + 9(1+t^2)v_\rho^4 + 6 t^2 v_\rho^2 v_\chi^2 \right] + \\
& \left.  + v_W^2\left[3(2t^4 - 6t^2 -9)v_\rho^2 -9(4t^2 -3)v_\chi^2\right] \right\}^{1/2}.
\end{split}
\end{equation}

\subsection{Solution}

The non-trivial solution in this case reads

\begin{equation} \label{eq:soldefrhn}
v_\rho^2=\frac{1-2 s_W^2}{2c_W^2}v_W^2.
\end{equation}

The analogous to Eq. (\ref{eq:tXsW}) now is 

\begin{equation} \label{eq:tXsWrhn}
t^2 \equiv \frac{g_X^2}{g^2} = \frac{s_W^2}{1-\frac{4}{3} s_W^2}.
\end{equation}

\subsection{Bounds}

The general expressions for the neutral current parameters of the model with right handed neutrinos are presented on \ref{app:couplingsrhn}, and the bound obtaining reasoning is identical to that presented last section. 

Now, the condition that picks out the right branch for the $Z$-mass and for every neutral current parameter is (see \ref{app:couplingssolrhn})

\begin{equation} \label{eq:boundrhn}
v_\chi > \frac{v_W}{\sqrt{2}c_W}\approx \SI{199}{GeV}
\end{equation} 

The masses of the charged vector bosons within this model are analytically the same as for the minimal model, but with $v_s=0$. For $U^{++}$, then, we have the bound

\begin{equation}
M_U=\sqrt{\frac{1}{4}g^2(v_\rho^2+v_\chi^2)} > \SI{78}{GeV}.
\end{equation}

Now the relation $v_\eta^2+v_\rho^2=(\SI{246}{GeV})^2$ sets $v_\eta\approx \SI{199}{GeV}$, with which we find, for $V^+$,

\begin{equation}
M_V=\sqrt{\frac{1}{4}g^2(v_\eta^2+v_\chi^2)} > \SI{89}{GeV}.
\end{equation}

The mass of the exotic neutral vector boson within the solution reads

\begin{equation}
M_{Z'}=\frac{g \,v_\chi^2}{4}\:\frac{4c_W^2 + (1-2s_W^2)\bar{v}_W^2}{4(3-7s_W^2+4s_W^4)}.
\end{equation}
Which presents the lower bound

\begin{equation}
M_{Z'}> \SI{89}{GeV}.
\end{equation}

\section{Conclusions}
\label{sec:con}

We have shown that Dias et al simple solution to the closing of the minimal 331 model at the electroweak scale actually implies a lower bound of $\SI{54}{GeV}$ on $v_\chi$ in the minimal model and of $\SI{59}{GeV}$ in the model with right-handed neutrinos -- which sets a lower bound for the five exotic vector boson masses. We have also added corrections to many expressions implied by the 331 minimal and version with right handed neutrinos' most general parametrization.

We recall that since $v_\chi$ is the scale of the breaking of the 331 into the Standard Model, it is theoretically predicted to be greater than $v_W\approx \SI{246}{GeV}$. However, if the 331 is viewed as an alternative (and not extension) of the SM, the projection of the representation content of the model onto the Standard Model is still possible, so that the relations presented here are still valid, and $v_\chi$ could, in principle, be smaller than $v_W$.

Finally, we note that current phenomenologic results such as presented in \cite{Barela:2019pmo}, which predicts $M_U \gtrsim \SI{1150}{GeV}$, or \cite{Binh:2021iww}, which derives the bound $M_{Z_2} \gtrsim \SI{3.1}{TeV}$, render our bounds useless in practice. The common phenomenologic impression that exotic particles' masses should be immensely greater than the bounds achieved in this paper, however, should be maintained carefully. This is because that type of search is mostly exploratory, in the sense that when studying a specific model or, specially, effecting model independent research, at least a few assumptions and benchmarks are eventually necessary. These simplifications might kill mechanisms that could allow for lower masses. For instance, in \cite{Barela:2019pmo} it is shown that for small enough values of a largely ignored matrix element, the bound  $M_U \gtrsim \SI{1150}{GeV}$ could be morphed to  $M_U \gtrsim \SI{200}{GeV}$.

The above discussion is not intended to argue that the bounds derived here might be useful, since they must be, even for the most skeptic of perspectives, indeed exaggeratedly low -- concerning masses possibly ruled out even by theoretical reasons such as perturbativity. The bounds, however, are interesting while being an \textit{exact} property of the theory and a needed remark concerning the work in \cite{Dias:2006ns}.

\section*{ACKNOWLEDGMENTS}
MB would like to thank CNPq for the financial support and, specially, Vicente Pleitez for many useful discussions.


\let\doi\relax


\begin{thebibliography}{99}

\bibitem{Dias:2006ns}
A.~G.~Dias, J.~C.~Montero and V.~Pleitez,
Phys. Rev. D \textbf{73}, 113004 (2006)
doi:10.1103/PhysRevD.73.113004
[arXiv:hep-ph/0605051 [hep-ph]].

\bibitem{Pisano:1991ee}
F.~Pisano and V.~Pleitez,
Phys. Rev. D \textbf{46}, 410-417 (1992)
doi:10.1103/PhysRevD.46.410
[arXiv:hep-ph/9206242 [hep-ph]].

\bibitem{Frampton:1992wt} 
P.~H.~Frampton,
\textrm{Chiral dilepton model and the flavor question},
Phys.\ Rev.\ Lett.\  {\bf 69}, 2889 (1992).

\bibitem{Foot:1992rh}
R.~Foot, O.~F.~Hernandez, F.~Pisano and V.~Pleitez,
Phys. Rev. D \textbf{47}, 4158-4161 (1993)
doi:10.1103/PhysRevD.47.4158
[arXiv:hep-ph/9207264 [hep-ph]].

\bibitem{Pleitez:1992xh}
V.~Pleitez and M.~D.~Tonasse,
Phys. Rev. D \textbf{48}, 2353-2355 (1993)
doi:10.1103/PhysRevD.48.2353
[arXiv:hep-ph/9301232 [hep-ph]].

\bibitem{Dias:2005yh}
A.~G.~Dias, C.~A.~de S.Pires and P.~S.~Rodrigues da Silva,
Phys. Lett. B \textbf{628}, 85-92 (2005)
doi:10.1016/j.physletb.2005.09.028
[arXiv:hep-ph/0508186 [hep-ph]].


\bibitem{Zyla:2020zbs}
P.~A.~Zyla \textit{et al.} [Particle Data Group],
PTEP \textbf{2020}, no.8, 083C01 (2020)
doi:10.1093/ptep/ptaa104

\bibitem{Barela:2019pmo}
M.~W.~Barela and V.~Pleitez,
Phys. Rev. D \textbf{101}, no.1, 015024 (2020)
doi:10.1103/PhysRevD.101.015024
[arXiv:1912.05900 [hep-ph]].

\bibitem{Binh:2021iww}
D.~T.~Binh, L.~T.~Hue, V.~H.~Binh, D.~V.~Soa and H.~N.~Long,
[arXiv:2109.08118 [hep-ph]].


\end{thebibliography}

\appendix

\section{Minimal model}

\subsection{General exact neutral current couplings}
\label{app:couplingsminimal}

These appendices exist to gather results for reference concerning neutral current parameters and do not add to the bound-obtaining reasoning itself.

Consider the neutral current lagrangian parametrized as usual \cite{Zyla:2020zbs}:

\begin{equation}
\mathcal{L}^{NC}=-\frac{g}{2c_W}\sum\limits_{i}\bar{\Psi}_i \gamma^{\mu} [(g^i_V-g^i_A \gamma_5)Z_{1 \mu}+(f^i_V-f^i_A \gamma_5)Z_{2 \mu}]\Psi_i.
\end{equation}

We calculate, independently, the values of the neutral current couplings following from the representation content of the minimal version of the 331 model. They are as follows (note that some values diverge from those presented in \cite{Dias:2006ns})

\begin{equation} \label{eq:firstpar}
g_V^\nu=g_A^\nu=N_1 \left(2m_2^2+\bar{v}_{\rho}^2-\frac{4}{3}\bar{v}_{W}^2-\frac{1}{3}\right);
\end{equation}

\begin{equation}
g_V^\ell= -c_W N_1\left(1 -\bar{v}_{\rho}^2\right);
\end{equation}

\begin{equation}
g_A^\ell=-c_W N_1\left(2m_2^2 -\frac{4}{3}\bar{v}_{W}^2+\bar{v}_{\rho}^2-\frac{1}{3}\right);
\end{equation}

\begin{equation}
g_V^{u_m}= c_W N_1 \left[m_2^2-\frac{2}{3}\bar{v}_{W}^2+\frac{1}{3}-\frac{2}{3}t^2\left(1 -\bar{v}_{\rho}^2\right)\right];
\end{equation}

\begin{equation}
g_A^{u_m}=c_W N_1 \left[m_2^2-\frac{2}{3}\bar{v}_{W}^2+\frac{1}{3}+\frac{6}{3}t^2\left(1 -\bar{v}_{\rho}^2\right)\right];
\end{equation}

\begin{equation}
g_V^{d_m}= c_W N_1 \left[-2m_2^2+\frac{4}{3}\bar{v}_{W}^2-\bar{v}_{\rho}^2+\frac{1}{3}+\frac{4}{3}t^2\left(1 -\bar{v}_{\rho}^2\right)\right];
\end{equation}

\begin{equation}
g_A^{d_m}= c_W N_1 \left[-2m_2^2+\frac{4}{3}\bar{v}_{W}^2-\bar{v}_{\rho}^2+\frac{1}{3}\right];
\end{equation}

\begin{equation}
g_V^{j_m}= c_W N_1 \left[m_2^2+\bar{v}_{\rho}^2-\frac{4}{3}\bar{v}_{W}^2-\frac{2}{3}+\frac{10}{3}t^2\left(1 -\bar{v}_{\rho}^2\right)\right];
\end{equation}

\begin{equation}
g_A^{j_m}= c_W N_1 \left[m_2^2+\bar{v}_{\rho}^2-\frac{4}{3}\bar{v}_{W}^2-\frac{2}{3}-\frac{6}{3}t^2\left(1 -\bar{v}_{\rho}^2\right)\right];
\end{equation}

\begin{equation}
g_V^{u_3}= -c_W N_1 \left[-2m_2^2-\bar{v}_{\rho}^2+\frac{4}{3}\bar{v}_{W}^2+\frac{1}{3}+\frac{8}{3}t^2\left(1 -\bar{v}_{\rho}^2\right)\right];
\end{equation}

\begin{equation}
g_A^{u_3}=-c_W N_1 \left[-2m_2^2-\bar{v}_{\rho}^2+\frac{4}{3}\bar{v}_{W}^2+\frac{1}{3}\right];
\end{equation}

\begin{equation}
g_V^{d_3}=-c_W N_1 \left[m_2^2-\frac{2}{3}\bar{v}_{W}^2+\frac{1}{3}+\frac{2}{3}t^2\left(1 -\bar{v}_{\rho}^2\right)\right];
\end{equation}

\begin{equation}
g_A^{d_3}=-c_W N_1 \left[m_2^2-\frac{2}{3}\bar{v}_{W}^2+\frac{1}{3}+\frac{6}{3}t^2\left(1 -\bar{v}_{\rho}^2\right)\right];
\end{equation}

\begin{equation}
g_V^{J}= -c_W N_1 \left[m_2^2+\bar{v}_{\rho}^2-\frac{4}{3}\bar{v}_{W}^2-\frac{2}{3}+\frac{14}{3}t^2\left(1 -\bar{v}_{\rho}^2\right)\right];
\end{equation}

\begin{equation} \label{eq:lastpar}
g_A^{J}=-c_W N_1 \left[m_2^2+\bar{v}_{\rho}^2-\frac{4}{3}\bar{v}_{W}^2-\frac{2}{3}-\frac{6}{3}t^2\left(1 -\bar{v}_{\rho}^2\right)\right].
\end{equation}

To get the $f$, analogous couplings to the $Z_2$, replace $N_1 \to N_2, N_2 \to N_1, m_1 \to m_2, m_2 \to m_1$.

In this appendix we have employed the abbreviations introduced by Dias:

 \begin{equation}\label{eq:massfactors}
\begin{split}
A&=\frac{1}{3}\left[ 3t^2\left(\bar{v}_\rho^2+1 \right) + \bar{v}_W^2 + 1\right], \\
R&=\left\lbrace 1-\frac{1}{3A^2}\left(4t^2+1 \right)\left[\bar{v}_W^2\left(\bar{v}_\rho^2+1 \right) -\bar{v}_\rho^4\right]\right\rbrace^{1/2},
\end{split}
\end{equation}

\begin{equation} \label{eq:neutralbmass}
\begin{split}
m_{1}^2&=\frac{2 M^2_{Z_1}}{g^2 v_\chi^2}=A(1-R), \\
m_{2}^2&=\frac{2 M^2_{Z_2}}{g^2 v_\chi^2}=A(1+R),
\end{split}
\end{equation}
where $M$ are the neutral vector boson masses and $\bar{v}_\alpha\equiv \frac{v_\alpha}{v_\chi}$. The normalization factors are

\begin{equation}\label{eq:normfactors}
\begin{split}
N_1^{-2}&=3 \left(2m_2^2+\bar{v}_\rho^2-\frac{4}{3}\bar{v}_W^2-\frac{1}{3}\right) + (\bar{v}_\rho^2-1)^2(4t^2+1), \\
N_2^{-2}&=3 \left(2m_1^2+\bar{v}_\rho^2-\frac{4}{3}\bar{v}_W^2-\frac{1}{3}\right) + (\bar{v}_\rho^2-1)^2(4t^2+1).
\end{split}
\end{equation}

\subsection{Neutral current couplings of the known fermions to the SM $Z$ within the closing solution}
\label{app:couplingssolminimal}

When constrained by the solution, the general values of the last section respective of the known fermions reduce to 

\begin{align}
g_V^\nu=g_A^\nu=
&\begin{cases}
\frac{1}{2}, & \text{if} \ v_\chi \geq \sqrt{\frac{1-4s_W^2}{2 c_W^2}}v_W \\
\frac{1}{2}\sqrt{\frac{1-4s_W^2}{3}}, & \text{otherwise}
\end{cases} \\
g_V^\ell=
&\begin{cases}
-\frac{1}{2}+2s_W^2, & \text{if} \ v_\chi \geq \sqrt{\frac{1-4s_W^2}{2 c_W^2}}v_W \\
\frac{1}{2}\sqrt{3(1-4s_W^2)}, & \text{otherwise}
\end{cases} \\
g_A^\ell=
&\begin{cases}
-\frac{1}{2}, & \text{if} \ v_\chi \geq \sqrt{\frac{1-4s_W^2}{2 c_W^2}}v_W \\
-\frac{1}{2}\sqrt{\frac{1-4s_W^2}{3}}, & \text{otherwise}
\end{cases} \\
g_V^{u_m}=
&\begin{cases}
-\frac{1}{2}+2s_W^2, & \text{if} \ v_\chi \geq \sqrt{\frac{1-4s_W^2}{2 c_W^2}}v_W \\
\frac{1}{2}\frac{-1+6s_W^2}{\sqrt{3(1-4s_W^2)}}, & \text{otherwise}
\end{cases} \\
g_A^{u_m}=
&\begin{cases}
\frac{1}{2}, & \text{if} \ v_\chi \geq \sqrt{\frac{1-4s_W^2}{2 c_W^2}}v_W \\
\frac{1}{2}\frac{-1-2s_W^2}{\sqrt{3(1-4s_W^2)}}, & \text{otherwise}
\end{cases} \\
g_V^{d_m}=
&\begin{cases}
\frac{1}{6}(-3+4s_W^2), & \text{if} \ v_\chi \geq \sqrt{\frac{1-4s_W^2}{2 c_W^2}}v_W \\
-\frac{1}{2}\frac{1}{\sqrt{3(1-4s_W^2)}}, & \text{otherwise}
\end{cases} \\
g_A^{d_m}=
&\begin{cases}
-\frac{1}{2}, & \text{if} \ v_\chi \geq \sqrt{\frac{1-4s_W^2}{2 c_W^2}}v_W \\
-\frac{1}{2}\sqrt{\frac{1-4s_W^2}{3}}, & \text{otherwise}
\end{cases} \\
g_V^{u_3}=
&\begin{cases}
\frac{1}{6}(3-8s_W^2), & \text{if} \ v_\chi \geq \sqrt{\frac{1-4s_W^2}{2 c_W^2}}v_W \\
\frac{1}{2}\frac{1+4s_W^2}{\sqrt{3(1-4s_W^2)}}, & \text{otherwise}
\end{cases} \\
g_A^{u_3}=
&\begin{cases}
\frac{1}{2}, & \text{if} \ v_\chi \geq \sqrt{\frac{1-4s_W^2}{2 c_W^2}}v_W \\
\frac{1}{2}\sqrt{\frac{1-4s_W^2}{3}}, & \text{otherwise}
\end{cases} \\
g_V^{d_3}=
&\begin{cases}
\frac{1}{6}(-3+4s_W^2), & \text{if} \ v_\chi \geq \sqrt{\frac{1-4s_W^2}{2 c_W^2}}v_W \\
-\frac{1}{2}\frac{1-2s_W^2}{\sqrt{3(1-4s_W^2)}}, & \text{otherwise}
\end{cases} \\
g_A^{d_3}=
&\begin{cases}
-\frac{1}{2}, & \text{if} \ v_\chi \geq \sqrt{\frac{1-4s_W^2}{2 c_W^2}}v_W \\
\frac{1}{2}\frac{1+4s_W^2}{\sqrt{3(1-4s_W^2)}}, & \text{otherwise}
\end{cases}
\end{align}
from which we may observe that the SM value is reproduced in every case when Eq. (\ref{eq:bound}) is satisfied.

\section{Model with right handed neutrinos}

\subsection{General exact neutral current couplings}
\label{app:couplingsrhn}

The general neutral current parameters within this model, introducing a new set of abbreviations, are found to be

\begin{equation} 
g_V^\nu=-\frac{1}{2}c_W N_1 \left(F_{31}+\sqrt{3}F_{81}\right);
\end{equation}

\begin{equation} 
g_A^\nu=-\frac{1}{6}c_W N_1 \left(3 F_{31}-\sqrt{3}F_{81}-4B_1 t\right);
\end{equation}

\begin{equation}
g_V^\ell=\frac{1}{6}c_W N_1 \left(3 F_{31}-\sqrt{3}F_{81}+8B_1 t\right);
\end{equation}

\begin{equation}
g_A^\ell=\frac{1}{6}c_W N_1 \left(3 F_{31}-\sqrt{3}F_{81}-4B_1 t\right);
\end{equation}

\begin{equation}
g_V^{u_m}= \frac{1}{6}c_W N_1 \left(-3 F_{31}+\sqrt{3}F_{81}-4B_1 t\right);
\end{equation}

\begin{equation}
g_A^{u_m}=\frac{1}{6}c_W N_1 \left(-3 F_{31}+\sqrt{3}F_{81}+4B_1 t\right);
\end{equation}

\begin{equation}
g_V^{d_m}= \frac{1}{6}c_W N_1 \left(3 F_{31}+\sqrt{3}F_{81}+2B_1 t\right);
\end{equation}

\begin{equation}
g_A^{d_m}= \frac{1}{6}c_W N_1 \left(3 F_{31}+\sqrt{3}F_{81}-2B_1 t\right);
\end{equation}

\begin{equation}
g_V^{D_m}= \frac{1}{3}c_W N_1 \left(-\sqrt{3}F_{81}+B_1 t\right);
\end{equation}

\begin{equation}
g_A^{D_m}= \frac{1}{3}c_W N_1 \left(-\sqrt{3}F_{81}-B_1 t\right);
\end{equation}

\begin{equation}
g_V^{u_3}= -\frac{1}{6}c_W N_1 \left(3F_{31}+\sqrt{3}F_{81}+6B_1 t\right);
\end{equation}

\begin{equation}
g_A^{u_3}=-\frac{1}{6}c_W N_1 \left(3F_{31}+\sqrt{3}F_{81}-2B_1 t\right);
\end{equation}

\begin{equation}
g_V^{d_3}=\frac{1}{6}c_W N_1 \left(3F_{31}-\sqrt{3}F_{81}\right);
\end{equation}

\begin{equation}
g_A^{d_3}=\frac{1}{6}c_W N_1 \left(3F_{31}-\sqrt{3}F_{81}-4B_1 t\right);
\end{equation}

\begin{equation}
g_V^{U}=\frac{1}{3}c_W N_1 \left(\sqrt{3}F_{81}-B_1 t\right);
\end{equation}

\begin{equation}
g_A^{U}=\frac{1}{3}c_W N_1 \left(\sqrt{3}F_{81}+B_1 t\right);
\end{equation}

For the values of the fermionic couplings with $Z_2$, make the replacements $N_1 \to N_2, F_{31} \to F_{32}, \; F_{81} \to F_{82}, B_{1} \to B_{2}$ in the above formulas, where the relevant symbols are defined by

\begin{equation}
\begin{split}
F_{31}&=3(3+t^2)\bar{v}_W^4+6(3+t^2)\bar{v}_\rho^4-3\bar{v}_W^2(3+R+t^2+6\bar{v}_\rho^2-t^2\bar{v}_\rho^2), \\
\frac{F_{81}}{\sqrt{3}}&=(3+t^2)\bar{v}_W^4-2(3+t^2-R)\bar{v}_\rho^2-\bar{v}_W^2[(-3+R+6\bar{v}_\rho^2+5t^2(1+\bar{v}_\rho^2)], \\
B_1&=-2t[(3+t^2)\bar{v}_W^4+\bar{v}_W^2(-6-R+t^2-6\bar{v}_\rho^2+4t^2\bar{v}_\rho^2)]+\bar{v}_\rho^2(-R+(3+t^2)(1+3\bar{v}_\rho^2)], \\
F_{32}&=-3(3+t^2)\bar{v}_W^4-6(3+t^2)\bar{v}_\rho^4-3\bar{v}_W^2[-3+R-6\bar{v}_\rho^2+t^2(-1+\bar{v}_\rho^2)], \\
\frac{F_{82}}{\sqrt{3}}&=(3+t^2)\bar{v}_W^4-2(3+t^2+R)\bar{v}_\rho^2-\bar{v}_W^2[(-3-R+6\bar{v}_\rho^2+5t^2(1+\bar{v}_\rho^2)], \\
B_2&=-2t[(3+t^2)\bar{v}_W^4+\bar{v}_W^2(-6+R+t^2-6\bar{v}_\rho^2+4t^2\bar{v}_\rho^2)]+\bar{v}_\rho^2(R+(3+t^2)(1+3\bar{v}_\rho^2)],
\end{split}
\end{equation}
where

\begin{equation}
\begin{split}
R=\{t^4(1+\bar{v}_W^2+3\bar{v}_\rho^2)^2+&9[1+\bar{v}_W^4+3\bar{v}_\rho^4-\bar{v}_W^2(1+3\bar{v}_\rho^2)]+ \\
+&6t^2[1+\bar{v}_W^4+3\bar{v}_\rho^2+6\bar{v}_\rho^4-\bar{v}_W^2(4+3\bar{v}_\rho^2)]\}^{1/2},
\end{split}
\end{equation}
with, still, $\bar{v}_\alpha\equiv \frac{v_\alpha}{v_\chi}$.

With this, the relation between symmetry and mass eigenstates is written

\begin{equation}
\begin{split}
W_3^\mu&=N_1 F_{31} Z_1^\mu + N_2 F_{32} Z_2^\mu + \frac{t}{\sqrt{1+\frac{4}{3}t^2}} A^\mu; \\
W_8^\mu&=N_1 F_{81} Z_1^\mu + N_2 F_{82} Z_2^\mu - \frac{t}{\sqrt{3+4t^2}} A^\mu; \\
B^\mu&=N_1 B_{1} Z_1^\mu + N_2 B_{2} Z_2^\mu + \frac{1}{\sqrt{1+\frac{4}{3}t^2}} A^\mu,
\end{split}
\end{equation}
where the normalization factors are as expected

\begin{equation}
\begin{split}
N_1=&\frac{1}{\sqrt{F_{31}^2+F_{81}^2+B_{1}^2}}, \\
N_2=&\frac{1}{\sqrt{F_{32}^2+F_{82}^2+B_{2}^2}}.
\end{split}
\end{equation}

\subsection{Neutral current couplings of the known fermions to the SM $Z$ within the closing solution}
\label{app:couplingssolrhn}

The SM $Z$ parameters simplified by the solution now are

\begin{align}
g_V^\nu=
&\begin{cases}
\frac{1}{2}, & \text{if} \ v_\chi > \frac{v_W}{\sqrt{2}c_W} \\
0, & \text{otherwise}
\end{cases} \\
g_A^\nu=
&\begin{cases}
\frac{1}{2}, & \text{if} \ v_\chi > \frac{v_W}{\sqrt{2}c_W} \\
0, & \text{otherwise}
\end{cases} \\
g_V^\ell=
&\begin{cases}
-\frac{1}{2}+2s_W^2, & \text{if} \ v_\chi > \frac{v_W}{\sqrt{2}c_W} \\
0, & \text{otherwise}
\end{cases} \\
g_A^\ell=
&\begin{cases}
-\frac{1}{2}+2s_W^2, & \text{if} \ v_\chi > \frac{v_W}{\sqrt{2}c_W} \\
0, & \text{otherwise}
\end{cases} \\
g_V^{u_m}=
&\begin{cases}
\frac{1}{6}(3-8 s_W^2), & \text{if} \ v_\chi > \frac{v_W}{\sqrt{2}c_W} \\
0, & \text{otherwise}
\end{cases} \\
g_A^{u_m}=
&\begin{cases}
\frac{1}{2}, & \text{if} \ v_\chi > \frac{v_W}{\sqrt{2}c_W} \\
0, & \text{otherwise}
\end{cases} \\
g_V^{d_m}=
&\begin{cases}
\frac{1}{6}(-3+4s_W^2), & \text{if} \ v_\chi > \frac{v_W}{\sqrt{2}c_W} \\
0, & \text{otherwise}
\end{cases} \\
g_A^{d_m}=
&\begin{cases}
-\frac{1}{2}, & \text{if} \ v_\chi > \frac{v_W}{\sqrt{2}c_W} \\
0, & \text{otherwise}
\end{cases} \\
g_V^{u_3}=
&\begin{cases}
\frac{1}{6}(3-8s_W^2), & \text{if} \ v_\chi > \frac{v_W}{\sqrt{2}c_W} \\
0, & \text{otherwise}
\end{cases} \\
g_A^{u_3}=
&\begin{cases}
\frac{1}{2}, & \text{if} \ v_\chi > \frac{v_W}{\sqrt{2}c_W} \\
0, & \text{otherwise}
\end{cases} \\
g_V^{d_3}=
&\begin{cases}
\frac{1}{6}(-3+4s_W^2), & \text{if} \ v_\chi > \frac{v_W}{\sqrt{2}c_W} \\
0, & \text{otherwise}
\end{cases} \\
g_A^{d_3}=
&\begin{cases}
-\frac{1}{2}, & \text{if} \ v_\chi > \frac{v_W}{\sqrt{2}c_W} \\
0, & \text{otherwise}
\end{cases}
\end{align}

\end{document}